\documentclass[%
 reprint,
 amsmath,amssymb,
 aps,
floatfix,
superscriptaddress
]{revtex4-1}
\usepackage{color}
\usepackage{graphicx}
\usepackage{dcolumn}
\usepackage{bm}
\usepackage{upgreek}
\usepackage{xfrac}
\newcommand{\textgreek}[1]{$\upmu$}

\begin{document}

\preprint{APS/123-QED}

\title{GHz-clocked teleportation of time-bin qubits with a telecom C-band quantum dot}


\author{M. Anderson}
\affiliation{Toshiba Research Europe Limited, 208 Science Park, Milton Road, Cambridge, CB4 0GZ, UK}
\affiliation{Cavendish Laboratory, University of Cambridge, JJ Thomson Avenue, Cambridge, CB3 0HE, UK}
\author{T. M\"{u}ller}
\email{tina.muller@crl.toshiba.co.uk}
\affiliation{Toshiba Research Europe Limited, 208 Science Park, Milton Road, Cambridge, CB4 0GZ, UK}
\author{J. Huwer}
\affiliation{Toshiba Research Europe Limited, 208 Science Park, Milton Road, Cambridge, CB4 0GZ, UK}
\author{J. Skiba-Szymanska}
\affiliation{Toshiba Research Europe Limited, 208 Science Park, Milton Road, Cambridge, CB4 0GZ, UK}
\author{ A.B. Krysa}
\affiliation{EPSRC National Epitaxy Facility, University of Sheffield, Sheffield, S1 3JD, UK}
\author{R.M. Stevenson}
\affiliation{Toshiba Research Europe Limited, 208 Science Park, Milton Road, Cambridge, CB4 0GZ, UK}
\author{J. Heffernan}
\affiliation{Departement of Electronic and Electrical Engineering, University of Sheffield, Sheffield, S1 3JD, UK}
\author{D.A. Ritchie}
\affiliation{Cavendish Laboratory, University of Cambridge, JJ Thomson Avenue, Cambridge, CB3 0HE, UK}
\author{A.J. Shields}
\affiliation{Toshiba Research Europe Limited, 208 Science Park, Milton Road, Cambridge, CB4 0GZ, UK}
\date{\today}

\begin{abstract}
\textbf{Abstract}
Teleportation is a fundamental concept of quantum mechanics with an important application in extending the range of quantum communication channels via quantum relay nodes. To be compatible with real-world technology such as secure quantum key distribution over fibre networks, such a relay node must operate at GHz clock rates and accept time-bin encoded qubits in the low-loss telecom band around 1550 nm. Here, we show that InAs/InP droplet epitaxy quantum dots with their sub-Poissonian emission near 1550 nm are ideally suited for the realisation of this technology. To create the necessary on-demand photon emission at GHz clock rates, we develop a flexible pulsed optical excitation scheme, and demonstrate that the fast driving conditions are compatible with a low multiphoton emission rate. We show further that, even under these driving conditions, photon pairs obtained from the biexciton cascade show an entanglement fidelity close to 90\%, comparable to the value obtained under cw excitation. Using asymetric Mach Zehnder interferometers and our photon source, we finally construct a time-bin qubit quantum relay able to receive and send time-bin encoded photons, and demonstrate mean teleportation fidelities of $0.82\pm0.01$, exceeding the classical limit by nearly 10 standard deviations.
\end{abstract}
\maketitle


\section{\label{sec:level1}Introduction}
Quantum networks can enable quantum technologies ranging from secure communication to distributed quantum computation using quantum nodes at separate locations \cite{Knill.2001, Kimble.2008}. Semiconductor quantum dots (QDs) are promising candidates to realise a variety of these technologies, where their atom-like energy structure and spin-photon interface \cite{Gao.2012} facilitate the entanglement of separate spins for distributed quantum computing \cite{Stockill.2017} or the generation of indistinguishable, ultra-pure single photons \cite{Somaschi.2016, Ding.2016} with strong polarisation entanglement \cite{Huber.2018} for photonic quantum network applications. Much progress has been made recently in improving these photonic qualities to create near-ideal, efficient, on-demand single and entangled photon sources \cite{Chen.2018, Liu.2019, Wang.2019b, Wang.2019}, even outperforming weak Poissonian sources in terms of single photon rates while avoiding the complications arising from multi-photon events \cite{Loredo.2017, He.2017}.

When establishing larger-scale quantum networks over existing fibre infrastructure, attenuation and loss of the weak quantum signal due to photon absorption is a major challenge, limiting the achievable length of quantum network links. In contrast to classical signals, the no-cloning theorem prevents the simple amplification of quantum states, leading to the proposal of quantum repeater schemes instead \cite{Briegel.1998}. A full quantum repeater requires either 2D photonic graph states \cite{Azuma.2015} or an operational quantum memory, of which practical demonstrations based on quantum dots are still outstanding despite recent progress in this direction \cite{Gangloff.2019,Denning.2019}. However, it is already feasible to operate quantum relay nodes \cite{Jacobs.2002} to increase the signal-to-noise ratio over extended distances. There are many proof-of-principle demonstrations of this concept \cite{Nilsson.2013, Varnava.2016, Huwer.2017, Reindl.2018}, however there are a few key differences in how QD technology and existing network applications such as quantum key distribution (QKD) are typically operated. These need to be overcome if practical integration of QD-based quantum relays with QKD systems is to be realised. They concern the operating wavelength, operating frequency and encoding scheme, which are all addressed in this work.

Importantly, for long-distance fibre quantum networks it is important to use the low-loss window in telecom fibres around 1550 nm (C-band) to maximise transmission of qubits. In contrast, the most mature QD devices operate at wavelengths around 900 nm. Encouragingly, there has been recent progress in developing low strain InAs QD devices with emission wavelength in the C-band directly using two alternative approaches, where one includes metamorphic buffers as relaxation layers in more traditional GaAs-based devices, and the other, the approach used here, consists in swapping the GaAs matrix material for InP. Both approaches have delivered single photon emission  \cite{Cade.2006, Benyoucef.2013, Paul.2017, Zeuner.2018} and small fine structure splitting (FSS) \cite{SkibaSzymanska.2017, Olbrich.2017}, resulting in entangled photon emission \cite{Olbrich.2017, Muller.2018}. However, compared to GaAs based C-band devices \cite{Nawrath.2019}, coherence times are an order of magnitude closer to the Fourier limit for InP-based devices \cite{Anderson.182019}, which remain the system of choice when implementing an interference-based application such as a quantum relay.

Further, QKD systems tend to use high clock rates of a GHz and above \cite{Boaron.2018}, while for QD systems clock rates under pulsed excitation are limited to the region of 80~MHz for C-band QDs, and even for lower wavelength QDs entangled photon pairs have not been generated at frequencies higher than 400 MHz \cite{Zhang.2015}.

Finally, long-distance QKD systems tend to use time-bin encoded qubits to increase robustness \cite{Brendel.1999, Dynes.2009} during long-distance fibre transmission. QDs on the other hand natively produce polarisation encoded single photons and entangled photons via the biexciton cascade. To overcome this incompatibility, schemes based on weak pulsed excitation have been developed to implement time-bin encoded single photons \cite{Lee.2018} and entangled photon pairs \cite{Simon.2005, Pathak.2011, Jayakumar.2014, Huber.2016} directly from quantum dots. Direct time bin entanglement generation from quantum dots has the advantage that larger FSSs are acceptable \cite{Simon.2005}, but this scheme cannot produce true on demand entangled pairs without the addition of a metastable state\cite{Simon.2005, Pathak.2011}, and it is technically demanding due to the need for phase control between the early and late photons, which can only be achieved with resonant two-photon excitation \cite{Jayakumar.2014, Huber.2016}. Alternatively, in the approach favoured here, qubit transcoder setups using asymmetric Mach Zehnder interferometers (AMZIs) can be used to create time-bin entangled photons from polarisation entangled ones \cite{Sanaka.2002, Versteegh.2015}, however demonstrations of this principle using C-band QDs are still outstanding.
\begin{figure}
\includegraphics[width=0.45\textwidth]{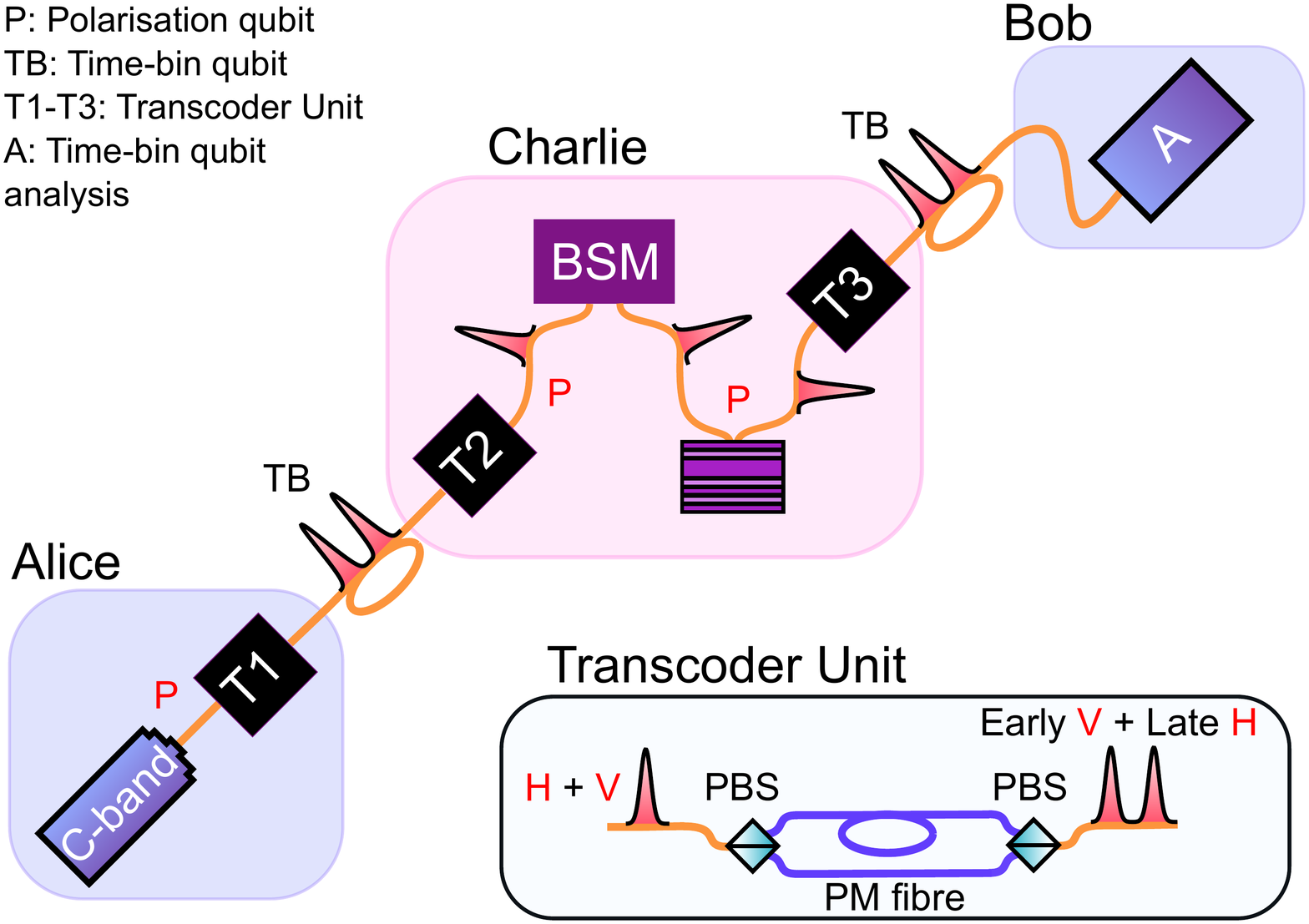} 
\caption{\textbf{Interfacing GHz-clocked time-bin qubits with a quantum dot relay. } Alice produces a GHz clocked, time-bin encoded qubit by sending polarisation encoded weak coherent pulses through an interferometric transcoding unit. Subsequent time encoded photons are then sent to a relay node Charlie where upon conversion back into polarisation, using a phase-stabilised second transcoding unit, the state can be teleported using a GHz-clocked polarisation entangled photon-pair source. After successful teleportation at Charlie, photons can be again transcoded to time bin before being sent on to a receiver Bob. }
\label{fig:concept}
\end{figure}
In this work, we simultaneously address the above three key challenges to create the first GHz-clock rate, on demand, time-bin compatible telecom wavelength quantum relay integrable with existing QKD technology and suitable for use with a variety of long-distance quantum network applications. A conceptual overview of our relay node is shown in Fig. \ref{fig:concept}. At the core of the relay station at Charlie is an InP-based InAs droplet epitaxy QD device, which offers low FSS and emission near 1550 nm with a high degree of polarisation entanglement \cite{Muller.2018} and long coherence times \cite{Anderson.182019}. The QDs are optically driven at GHz clock rates using a flexible pulse generation setup developed in-house. To interface the time-bin entangled qubits at Charlie's input with the emitted polarisation-entangled photon pairs, we construct a qubit transcoding unit based on stabilised AMZIs as described further below. Such a transcoder is also used to convert Charlie's output back to efficient time-bin encoding before being sent over a fibre link to Bob, and to produce time-bin encoded qubits at Alice to send to Charlie from GHz clocked, polarisation encoded pulses. In this manner we are able to create and teleport arbitrary time-bin encoded superpositions of the basis states.\\
\indent The paper is organised as follows: We start by characterising the QD emission under GHz clocked pulsed optical excitation, measuring lifetimes and autocorrelations to reveal the absence of multiple photons in the same emission cycle.  We then measure the entanglement fidelity achieved under these driving conditions using full tomography and compare it to the entanglement fidelity obtained under continuous-wave (CW) and low repetition rate (100 MHz) excitation. Next, we characterise the teleportation process with time-bin encoded states in three independent bases, before demonstrating that a time-bin encoded polar state is indeed mapped to the correct output time bin at Bob.

\section{GHz-clocked emission}

\begin{figure*}
\includegraphics[width=\textwidth]{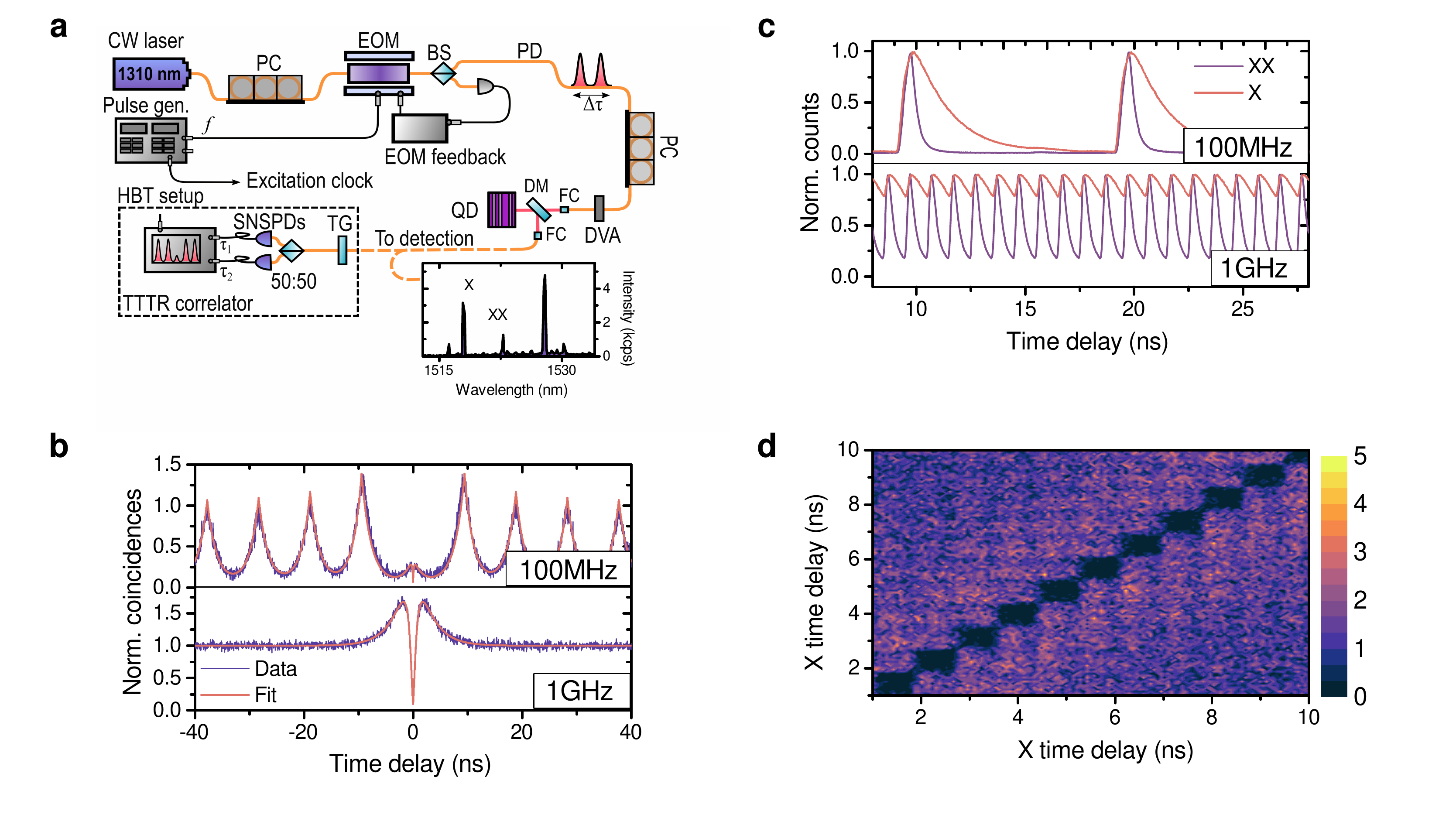} 
\caption{\textbf{GHz-clocked emission from a QD emitting near the telecom C-band} \textbf{a} Schematic of the excitation setup. A 1310~nm CW laser is modulated by an EOM with active voltage stabilisation in order to create flexible excitation pulses. GHz clocked emission is collected from the QD and sent to the detection system where time-resolved measurements can be made using the excitation clock from the pulse generator. PC: polarisation controller, EOM: Electro-Optic modulator, BS: beamsplitter, PD: photodiode, DM: dichroic mirror, FC: fibre coupler, DVA: digital variable attenuator, TG: transmission grating, SNSPDs: superconducting nanowire single photon detectors. Inset: spectrum of the QD measured under pulsed excitation at a power where the exciton intensity begins to saturate. The emission lines of the exciton (X) and the biexciton (XX) are labelled. \textbf{b} Second-order intensity correlations of the X for 100~MHz (top) and 1.07~GHz (bottom) excitation frequencies. Solid lines show fits to the data, where here the 1.07~GHz data appears completely CW. \textbf{c} Time-resolved intensity of the X and the XX emission lines at a repetition frequency of 100MHz (top) and 1.07~GHz (bottom). \textbf{d} Second-order correlation of the X photons, measured with respect to the excitation laser clock. The data binned on a 40x40 ps grid shows a leading diagonal of empty photon counts corresponding to the low probability of finding two X photons emitted in the same clock-cycle. Colour-bar denotes normalised coincidences.  }
\label{fig:emission}
\end{figure*}

We begin by introducing our scheme for creating flexible excitation pulses using an electro-optic modulator (EOM), a pulse generator and a DC offset to precisely control the polarisations created. A schematic of the setup used is shown in  Fig. \ref{fig:emission}(a), where a 1310-nm CW laser is sent through a 20-GHz bandwidth EOM and a polariser for intensity modulation. A pulse generator is used to define narrow (FWHM of 160 ps) excitation pulses. In order to maintain $>$20~dB suppression of the laser between the excitation pulses, we measure the average beam intensity and use a feedback control loop to the DC offset of the EOM to keep it constant. The resulting excitation pulses are then passed through a fibre polarisation controller (PC) and a digital variable attenuator (DVA) before being sent to the QD. With this setup, we are able to achieve a high level of control over the excitation conditions at various frequencies of interest.

To characterise the behaviour of our source under pulsed excitation, we examine the time-resolved response of the QD. A spectrum of the selected QD emitting near the telecom C-band when excited at a repetition rate of 1.07~GHz can be seen in Fig. \ref{fig:emission}(b), where several emission lines originating from different excitonic configurations can be seen. The lines of interest for this work are the neutral biexciton (XX) and exciton (X), which have been determined by polarisation resolved and intensity correlation measurements. The fine structure splitting (FSS) is determined to be $6.2 \pm 0.1~\upmu$eV. We first measure the time-resolved intensity for  a repetition rate of $f$=100~MHz and compare it to the time-resolved intensity at $f$=1~GHz. Both curves are shown in Fig. \ref{fig:emission}(c) for the X and XX transitions. The lifetime measurements show the expected behaviour with a sharp turn-on and a relatively short lifetime of $256 \pm 6$~ps for the XX at both repetition rates, and a slower turn-on due to filling effects and longer lifetime of $1.56 \pm 0.01$~ns at both repetition rates for X. At a 100-MHz repetition rate, this leads to well-separated pulses of QD emission.  For the above-GHz excitation rate common for many QKD protocols however, the excitation period becomes comparable to the radiative lifetimes, reducing the on-off contrast of the resulting emission to around 12$\%$ for the X and 69$\%$ for the XX.

One might expect that the overlapping of radiative decay cycles will affect the degree of multiphoton suppression under high repetition rate driving. Looking at the intensity autocorrelations using the Hanbury-Brown and Twiss (HBT) setup shown in Fig. \ref{fig:emission}(a), in a standard histogram of the delays between photons, we find the expected pulse signature when driving at 100 MHz, as shown in Fig. \ref{fig:emission}(d), where the finite intensity for the zero-delay pulse is mainly due to re-excitation effects. For a 1~GHz repetition rate however we see a strong CW-like signature, including bunching wings around the zero-delay dip. Interestingly though, when comparing the excitation repetition rates of 100~MHz and 1~GHz, we find the degree of multiphoton suppression is comparable with $g^{2}(0)$ values of $0.14 \pm 0.01$ and $0.09 \pm 0.05$, respectively, determined by fits to the experimental data. In both cases, the $g^{2}(0)$ is far below the 0.5 threshold required to prove single photon emission. For the 100~MHz data, the model includes the effects from both a shelving state (increased probability for time delays $>\vert\tau_{0}\vert$) and re-excitation (small volcano-like feature around the zero delay \cite{Jons.2017} ). For the 1.07-GHz measurement, a standard multi-level CW model is used.

To recover the signature of true on-demand, pulsed single photon emission even at GHz-clock rate driving conditions, we measure the arrival times of both X photons in the HBT setup with respect to a common clock. The result of this measurement can be seen in Fig. \ref{fig:emission}(e). Remarkably, we find that along the diagonal, corresponding to zero-delay coincidences of both X detectors, there is an absence of photons within the same 0.935x0.935~ns square determined by the 1.07-GHz repetition rate, with a $g^{2}[0] = 0.33 \pm 0.01$ . This shows clearly that there is a lack of multi-photon events within one entire excitation and emission period. Hence, despite the long radiative lifetime of this QD, we are able to generate single-photon emission clocked at 1~GHz.

\begin{figure}
\includegraphics[width=0.5\textwidth]{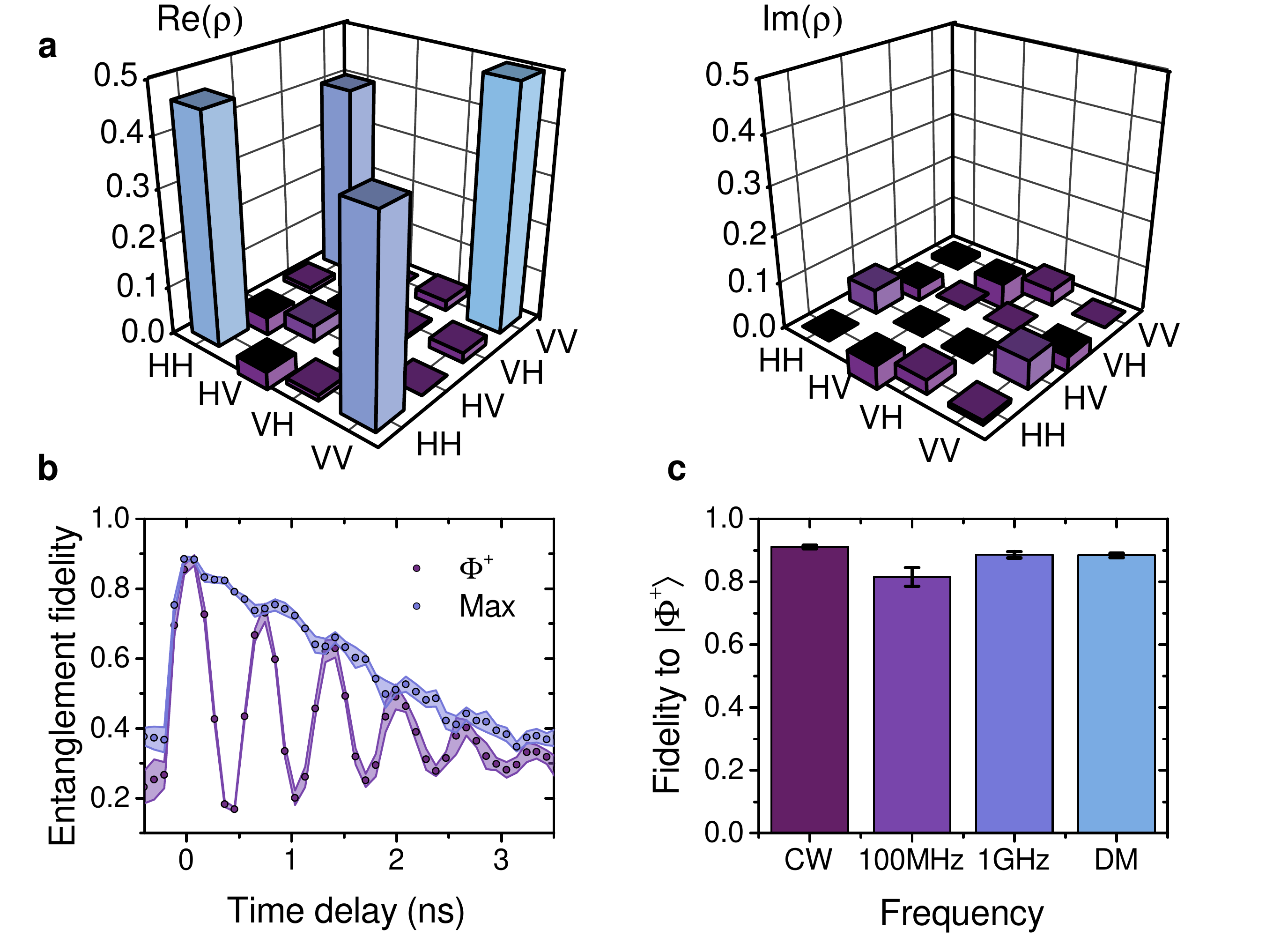}
\caption{\textbf{Degree of entanglement under GHz-clocked excitation}. \textbf{a} Real (left) and imaginary (right) part of the two-photon density matrix, reconstructed using a time window of 96~ps, under pulsed optical excitation clocked at 1.07~GHz. \textbf{b} Evolution of the fidelity of the two-photon state to the $\Phi^{+}$ Bell state and a maximally entangled state, calculated from the two-photon density matrix for different 96~ps time windows. The maximally entangled state follows that of the time-evolving state typically seen in our work \cite{Muller.2018}. \textbf{c} Comparison of the maximum fidelity to the $\Phi^{+}$ Bell state for different driving frequencies, see text for details.}
\label{fig:ent}
\end{figure}

The next consideration when preparing for pulsed teleportation is to look at the effect of the pulse repetition rate on the entanglement generated by the XX-X cascaded emission. In this process, after excitation of the XX, subsequent radiative recombination to the intermediate X state and then to the ground state results in a polarisation entangled two-photon state of the form

\begin{equation}\label{ent}
{|\Psi(\tau)\rangle = \frac{1}{\sqrt{2}}\left[ |H_{XX}H_{X}\rangle + e^{i\Delta E \tau/\hbar} |V_{XX}V_{X}\rangle \right ]},
\end{equation}

where $\Delta E$ is the FSS. In the limit of vanishing FSS, the state is no longer time-evolving and has the form of the symmetric $\vert \Phi^{+} \rangle$ Bell state. For excitation at 1~GHz we fully characterise this state via quantum state tomography \cite{Michler.2009,James.2001} to reconstruct the two-photon density matrix from a complete set of XX-X polarisation cross-correlations measured in three mutually unbiased bases. We utilise the maximum-likelihood approach \cite{James.2001} in reconstruction to ensure that the resulting matrix is Hermitian and positive semi-definite. The resulting density matrix expressed in the QD eigenbasis is shown in Fig. \ref{fig:ent}(a) for a 96~ps window around the zero delay, and has a strong fidelity to the  $\vert \Phi^{+} \rangle$ Bell state of $0.88\pm0.01$. The second strongest contribution, at 0.04, are classical correlations between H and V where the phase information has been lost. These are most likely a consequence of the finite timing resolution of our detectors ($\sim$100 ps). Further deviations of the matrix away from the ideal Bell state can be attributed to experimental imperfections as well as contributions from uncorrelated emission from the sample. From the density matrix we can also evaluate the concurrence, with the maximum determined here to be $0.80\pm0.01$ and thus proving a strong coherence in the entangled state.

As a result of the FSS, a time-dependant phase is accrued during the time spent in the intermediate X state, as described in Equation \ref{ent}. Thus, the expected fidelity to the Bell state oscillates at a beat frequency proportional to the magnitude of the FSS. We evaluate the fidelity of the reconstructed matrix for multiple time windows of 96~ps in order to reveal these quantum oscillations. The resulting fidelity is plotted in Fig. \ref{fig:ent}(b) alongside the fidelity to a maximally entangled state, which follows the fidelity to the time-evolving state of Equation \ref{ent} that is usually displayed in our work \cite{Muller.2018}.

The entanglement fidelity for the different driving frequencies considered here is compared in Fig. \ref{fig:ent}(c). We see that the values compare very well to those achieved under CW excitation ($0.91\pm0.01$), 100~MHz ($0.82\pm0.03$) and 1 GHz ($0.89\pm0.01$) measured using the standard approach \cite{Muller.2018}, where the latter agrees perfectly with the value extracted from the density matrix (shown for completeness).

\section{\label{sec:level2}Time-bin encoded quantum\protect\\ relay}

\begin{figure*}
\includegraphics[width=\textwidth]{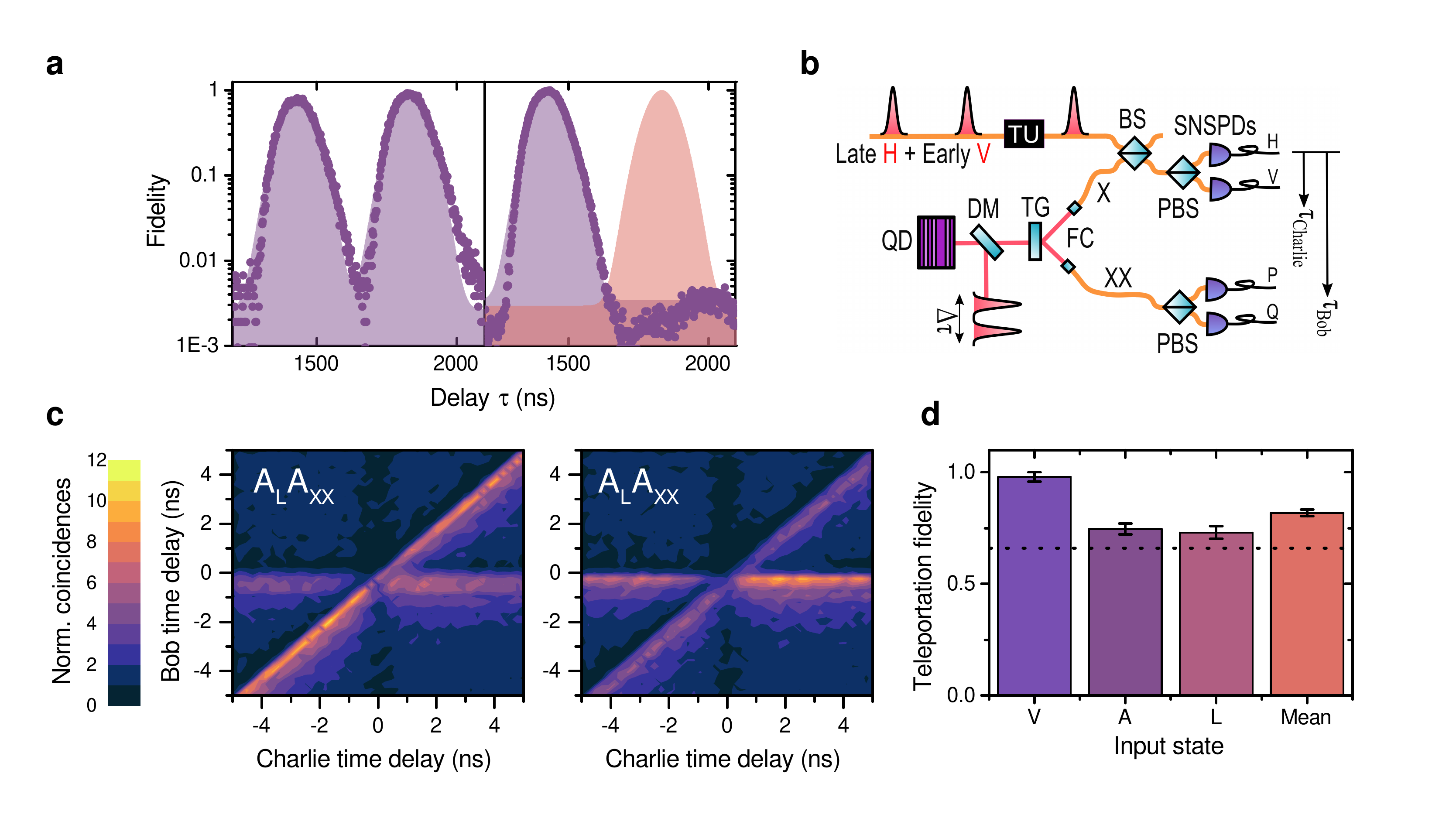}
\caption{\textbf{Superposition basis teleportation at 1~GHz}. \textbf{a} Input laser qubits in the time-bin basis for a superposition state (left) and an early state (right). \textbf{b} Schematic of the polarisation basis relay station where a time-bin input laser qubit is sent through a transcoder unit (TU) before being teleported using an entangled photon pair generated from the QD. \textbf{c} Normalised third-order correlations for sending input laser qubits in a superposition of $\vert e\rangle$ and $\vert l\rangle$ time bins, which is subsequently mapped to an $\vert A\rangle$ polarized photon, and measuring $\vert A\rangle$ (left) and $\vert D\rangle$ (right) polarized photons at Bob. \textbf{d} Teleportation fidelities for a complete set of orthogonal input states, calculated from the third-order correlations centered on the time $t_{Bob} = t_{Charlie} = 0$ , the point where all three photons are measured simultaneously. Results for the most statistically significant equivalent post-selection window size of 228~ps are shown. Each state surpasses the classical threshold of $\sfrac{2}{3}$ and a mean fidelity of $0.82 \pm 0.01$ beats this limit by more than $10\sigma$.}
\label{fig:teleport}
\end{figure*}

With the GHz-clocked emission and entanglement generation characterised, we proceed by performing teleportation of time-bin qubits. For the generation of the input qubits at Alice, we use a similar setup to that described in Fig. \ref{fig:emission}(a), where a feedback controlled EOM is used to create GHz-clocked pulses with a width of 130~ps (FWHM) from a tuneable telecom C-band CW laser. These pulses are then encoded using a qubit transcoder (as described in Fig. \ref{fig:concept}), before being sent to the relay station at Charlie. In this way, logical time-bin encoded states are created by mapping $V$ polarised photons onto the early time bin, and $H$ polarised photons onto the late time bin. A phase can then be encoded between the early and the late pulses by fine-tuning the delay between the two arms in order to create superposition basis states. Examples of the simulated laser input states are shown in Fig. \ref{fig:teleport}(a) for both a superposition basis state and an early logical state. To interface these qubits with the QD polarisation relay station at Charlie, a phase-matched decoding interferometer is used to map the states back into polarisation, allowing the teleportation to proceed. The setup we use for the teleportation is similar to that described in our previous work \cite{Anderson.182019}, and the main experimental features are shown in Fig. \ref{fig:teleport}(b) for clarity. An input qubit containing the polarisation encoded information to be teleported is incident upon a beam splitter together with the X photon from the QD entangled photon pair. When the photons arriving at the beam splitter are indistinguishable, they will exit through the same port as a result of the Hong-Ou-Mandel (HOM) interference effect. A polarising beam splitter (PBS) placed after the interfering beam splitter is then used to perform a Bell state measurement (BSM) which heralds the teleportation of the quantum state to the XX photon, measured at a second PBS.

In order to characterise the teleportation, we measure the triple coincidences $HVP$ and $HVQ$, corresponding to the successful detection of the BSM at Charlie, and a $P$ or $Q$ polarised photon at Bob, respectively. An example of such a third-order correlation is shown in Fig. \ref{fig:teleport}(c) for the case of teleporting a superposition basis state. Here, the input laser is prepared in an equal superposition of $\vert e\rangle$ and $\vert l\rangle$. This state is then decoded into polarisation and aligned to the QD $\vert A\rangle$ state. Three-photon coincidences $HVA$ and $HVD$ are then measured where, due to the experimental setup, the effect of this teleportation protocol is to map an $\vert A\rangle$ input to an $\vert A\rangle$ output at Bob. The peak in triple coincidences at $t_{Bob} = t_{Charlie} = 0$ is therefore expected to be seen when examining the $HVA$ correlation, and this is indeed the case as shown in Fig. \ref{fig:teleport}(c). Conversely, we see that there is an absence of three-photon events when looking at $HVD$ correlations, corresponding to a teleportation fidelity  $f^{T}_{A}=g^{(3)}_{A}(0)/(g^{(3)}_{A}(0)+g^{(3)}_{D}(0))$ of $0.75 \pm 0.03$ for the most significant post-selection window size of 228~ps. Measurement of at least three mutually orthogonal input states is required in order to prove the quantum nature of the teleportation \cite{Metcalf.2014} and so we measure the relevant correlations for the logical state $\vert e\rangle$, subsequently mapped to $\vert V\rangle$, as well as another equal superposition state subsequently mapped to $\vert L\rangle$. The resulting fidelities are shown in Fig. \ref{fig:teleport}(d), where we achieve a mean teleportation fidelity of $0.82 \pm 0.01$, more than 10 $\sigma$ beyond the classical limit of $2/3$. At the expense of teleported photons, a maximum mean fidelity of $0.92\pm0.04$ can be achieved for a much smaller post-selection window of 85~ps. These fidelities are comparable to those reached for teleportation of polarisation qubits \cite{Anderson.182019}, proving the accuracy and stability of the time bin to polarisation conversion interferometers.

We note that time gating is routinely employed in QKD experiments operating at such frequencies and indeed, this time gating and the resultant avoidance of detector dark counts when no teleported photons are expected are responsible for the favourable scaling of the signal-to-noise ratio with distance for a quantum relay compared to direct transmission \cite{Jacobs.2002}. Of course, a reduction of both the radiative lifetime and the FSS of our dots would result in more photons of higher fidelity within the time window and thus would improve the efficiency of the quantum relay, while relaxing the requirements of the exact width of the gate used.


\begin{figure}
\includegraphics[width=0.5\textwidth]{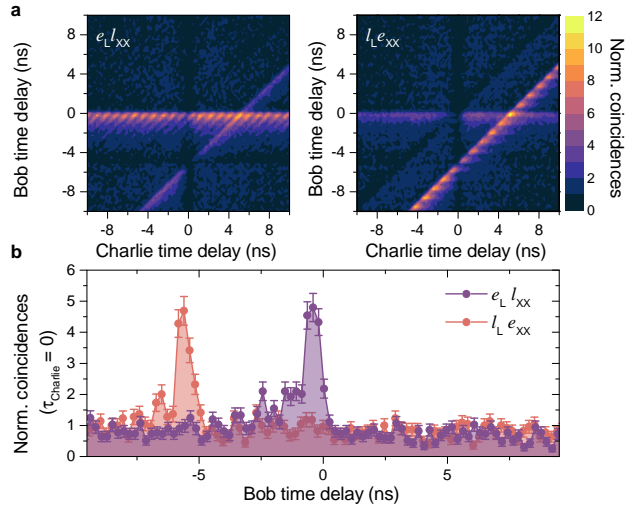} 
\caption{\textbf{Time-bin logical basis teleportation}. \textbf{a} Normalised third-order correlations for sending input laser qubits in $\vert e\rangle$ and $\vert l\rangle$ time bins, which are subsequently mapped to an $\vert l\rangle$ and $\vert e \rangle$ time bins after sending Bob photons back through the interferometer. \textbf{b} Teleportation fidelities along $t_{Charlie} = 0$ showing that input $\vert e\rangle$ is mapped to $\vert l\rangle$, corresponding to the polarization mapping of $\vert H\rangle$ to $\vert V\rangle$ in the teleportation.}
\label{fig:timebin}
\end{figure}

It is straightforward to now show that we can also convert the teleported photon back to time-bin encoding at Charlie before sending it to Bob. As a proof-of-principle, we teleport logical basis states, $\vert e\rangle$ and $\vert l\rangle$, except this time the output photon is then encoded back into the time-bin frame using a duplicate qubit transcoder. We implement this by sending the Bob photon back through the decoding interferometer, which guarantees a stable delay between early and late pulses for the two interferometers.

To determine that the output photon has been successfully teleported, we measure three-photon coincidences correlating a successful BSM in the polarisation basis with the time-bin arrival of the XX photons, for a given time-bin input laser qubit $\vert e\rangle$ or $\vert l\rangle$. First we consider the case when teleporting the $\vert e\rangle$ input laser qubit. Given that after qubit transcoding, the teleporter will map $H$ polarised photons to $V$ and vice versa, we expect that the teleporter will map the early input photons to late output photons. Indeed, this can be seen in Fig \ref{fig:timebin}(a), where we see a peak of coincidences at the time $t_{Bob} = t_{Charlie} = 0$, and an absence at $t_{Bob} = -5$~ns along the $ t_{Charlie} = 0$ axis, confirming that Bob's photon went through the long arm of the conversion interferometer. The other characteristic feature of these plots is the absence of coincidences along the $t_{Bob} = t_{Charlie}$ axis, corresponding to simultaneous measurement of the laser/X and XX photons. These events have also been shifted by 5~ns to $t_{Bob} = t_{Charlie} -5$~ns, where we see peaks of coincidences at the 1.07~GHz frequency of the system clock. Similarly, when sending a late photon, the expected outcome is an early one, as confirmed in the right-hand plot of Fig \ref{fig:timebin}(b). Here, a distinct peak of coincidences is seen at  $t_{Bob} = t_{Charlie} = -5$, which corresponds to a successful mapping of the late to early time bins. The arrival time of early and late teleported photons are highlighted in Fig \ref{fig:timebin}(b) where we plot a cut along the $t_{Charlie} = 0$ axis. The peak, and absence, of photons in the correct time bins results in a post-selected teleportation fidelity of $0.89\pm0.04$ for both logical time-bin encoded input states. This fidelity is comparable to that reached without the time-bin conversion for Bob's photon, confirming the accuracy of the conversion process.


\section{\label{sec:level4}Conclusion}

In conclusion, we have successfully addressed three of the major hurdles for integrating quantum dot devices with long distance quantum networks. We have shown GHz-clocked emission from a QD emitting near the telecom C-band while maintaining a single-photon level of $0.09 \pm 0.05$ post-selected and $0.33 \pm 0.01 $ across the entire excitation cycle, and recording the highest fidelity and concurrence of entanglement generated non-resonantly from the cascaded emission of the XX for this type of source. The GHz-clocked entangled photon-pair source was then used to interface with a time-bin encoded laser qubit to perform a pulsed quantum relay at telecom wavelengths.

While even a very inefficient quantum relay can beat the signal to noise ratio of a direct quantum link at long enough distances \cite{Jacobs.2002}, improvements to the photon extraction efficiency of our device would be desirable for practical applications. Including our device into recently demonstrated structures suitable for entangled photon extraction such as a circular Bragg grating \cite{Liu.2019, Wang.2019} or a broadband antenna \cite{Chen.2018} could boost the probability of extracting an entangled photon pair by four orders of magnitude. At the same time, such devices could lead to much shortened radiative lifetimes through Purcell enhancememnt of the emission \cite{Gerard.1998}, which would not only allow for even higher clock rates and shorter delays between the $\vert e\rangle$ and $\vert l\rangle$, but could also bring the emitted photons closer to the Fourier transform limit \cite{Bennett.2016} and relax time-gating or postselection criteria \cite{Reindl.2018} for the quantum relay. We expect these straightforward improvements to lead to the realisation of a true push-button quantum teleportation scheme easily integrable into existing quantum technology in the telecom C-band.

\begin{acknowledgments}
The authors acknowledge partial financial support from the Engineering and Physical Sciences Research Council and the UK's innovation agency, Innovate UK. M. A. gratefully acknowledges support from the Industrial CASE award funded by the EPSRC and Toshiba Research Europe Limited.\\


\end{acknowledgments}


%


\end{document}